\documentclass[aps,prb,showpacs,citeautoscript,superscriptaddress,twocolumn]{revtex4-1}

\usepackage{graphicx}
\usepackage{amsmath}
\usepackage{amssymb}
\usepackage{color}
\usepackage{subfigure}

\newcommand{\bea}{\begin{eqnarray*}}
\newcommand{\eea}{\end{eqnarray*}}
\newcommand{\bne}{\begin{equation*}}
\newcommand{\ede}{\end{equation*}}

\newcommand{\bnen}{\begin{equation}}
\newcommand{\eden}{\end{equation}}
\newcommand{\bean}{\begin{eqnarray}}
\newcommand{\eean}{\end{eqnarray}}
\newcommand{\bnsn}{\begin{subequations}}
\newcommand{\edsn}{\end{subequations}}

\newcommand{\bna}{\begin{array}}
\newcommand{\eda}{\end{array}}
\newcommand{\bnm}{\begin{enumerate}}
\newcommand{\edm}{\end{enumerate}}
\newcommand{\bni}{\begin{itemize}}
\newcommand{\edi}{\end{itemize}}

\renewcommand{\vec}[1]{\text{\boldmath{$ #1 $}}}

\newcommand{\ket}[1]{| #1 \rangle}
\newcommand{\bra}[1]{\langle #1 |}

\begin{document}
%\title{Spin-valley qubits in bent carbon nanotubes: \\ the effect 
%of inhomogeneous valley mixing (v6)}
%\title{Shape-sensitive Pauli blockade in bent carbon nanotubes}
\title{Coulomb-blockade and Pauli-blockade magnetometry}
%alternative: 
% Quantum-dot magnetometry 

\author{G\'abor Sz\'echenyi}
\affiliation{Institute of Physics, E\"otv\"os University, Budapest, Hungary}

\author{Andr\'as P\'alyi}
\affiliation{Institute of Physics, E\"otv\"os University, Budapest, Hungary}
\affiliation{MTA-BME Condensed Matter Research Group
and Department of Physics,
Budapest University of Technology and Economics, Budapest, Hungary}

\date{\today}

\begin{abstract}
Scanning-probe magnetometry is
a valuable experimental tool to investigate magnetic phenomena at the 
micro- and nanoscale. We theoretically analyze the possibility of 
measuring magnetic fields via  
the electrical current flowing through quantum dots.
We characterize the shot-noise-limited
magnetic-field sensitivity of 
two devices: a single dot in the Coulomb blockade regime, 
and a double  dot in the Pauli blockade regime. 
Constructing such magnetometers using
carbon nanotube quantum dots would benefit from the 
large, strongly anisotropic and controllable g tensors, 
the low abundance of nuclear spins, 
and the small detection volume allowing for
nanoscale spatial resolution; we estimate that
a sensitivity below $1 \, \mu\text{T}/\sqrt{\text{Hz}}$ can
be achieved with this material. 
As quantum dots have already proven to be useful 
as scanning-probe electrometers, our proposal
highlights their potential as hybrid 
sensors having in situ switching capability between electrical 
and magnetic sensing. 
\end{abstract}


%07.55.Ge Magnetometers for magnetic field measurements
%73.63.Kv Quantum dots 
%73.63.Fg Nanotubes 
%73.23.Hk Coulomb blockade; single-electron tunneling

\narrowtext\maketitle

\section{Introduction}
The detection of weak magnetic fields with high spatial resolution is a task of great importance in diverse areas, from fundamental physics and chemistry to practical applications in data storage and medical imaging. This task can be tackled by scanning-probe magnetic-field sensors,
based on various operating principles\cite{Bending,Martin_mfm,Chang_hall,Kirtley_apl,PhysRevLett.89.130801,Rugar2004,Taylor2008,Milde_skyrmion,Switkes,PhysRevLett.107.130801}.
%such as superconducting quantum interference devices, Hall probes 
%\cite{Bending}, atom-cloud magnetometers \cite{PhysRevLett.
%89.130801}, magnetically functionalised mechanical resonators 
%\cite{Rugar2004},\cite {PhysRevLett.107.130801}, or spin qubit 
%magnetometers \cite{Taylor2008}.
Low-temperature scanning-probe magnetometry has
been successfully used to image a range of 
nontrivial magnetic phenomena, e.g.,
vortices in superconductors\cite{Kirtley_vortices,Pelliccione,Thiel}, 
exotic magnetic structures\cite{Milde_skyrmion,Kezsmarki_skyrmion,RenshawWang}, 
and current-induced magnetic fields in various 
systems\cite{Bluhm_persistentcurrent,Kalisky-laosto} 
including 
topological insulators\cite{Spanton,Nowack_hgte}.
The applicability of the different magnetic-field sensors 
 (SQUIDs, Hall bars, NV centers, etc)
for specific tasks is determined by a number of 
characteristics, including magnetic-field sensitivity and
detection volume, the latter one related to the achievable
spatial resolution. 
%Key characteristics of these sensors, determining their applicability for 
%specific tasks, include magnetic-field sensitivity, spatial resolution, 
%magnetic-field range of operation, temperature range of operation, 
%frequency range of operation. 
A key challenge is to improve 
the capabilities of these sensors, either by advancing existing designs, 
or by devising completely new principles and devices.

\begin{figure*}
\begin{center}
\includegraphics[width=1.6\columnwidth]{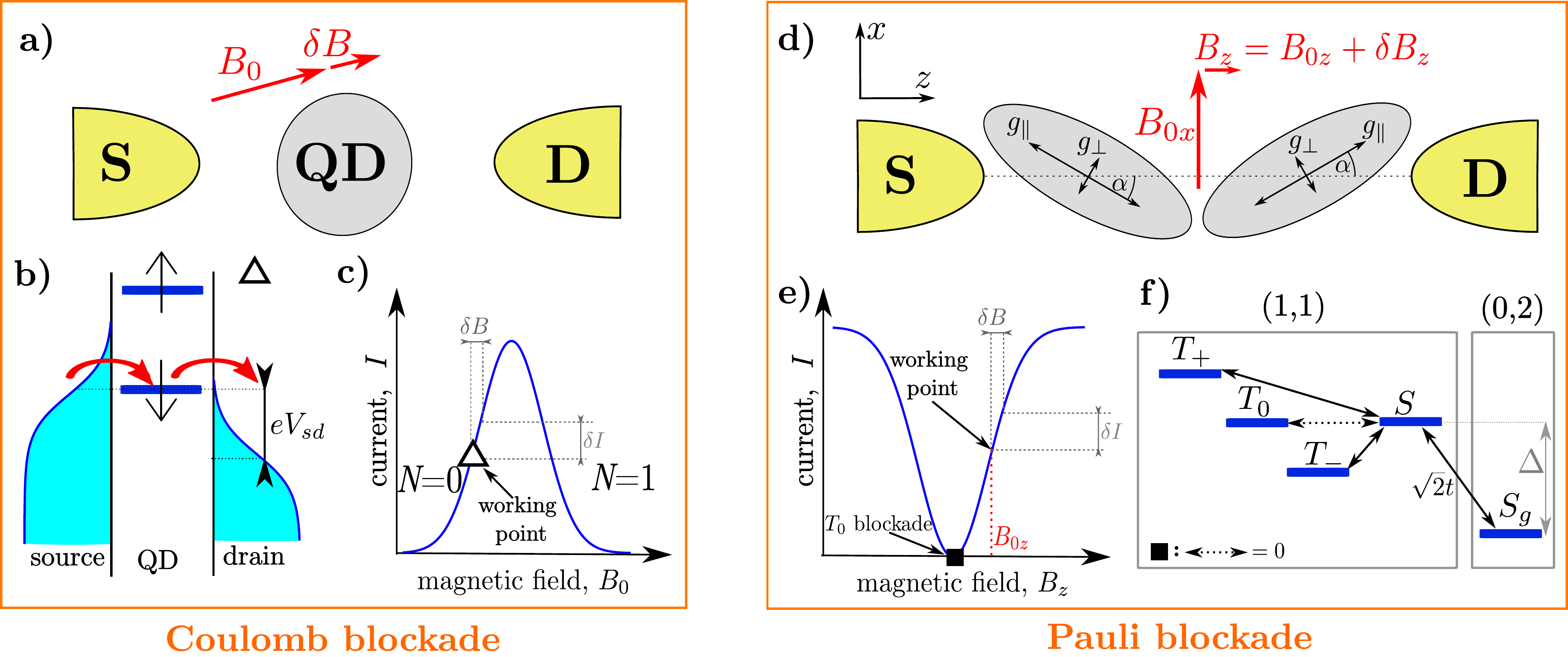}
\end{center}
%\vspace{-0.4cm}
\caption{(Color online)
Coulomb-blockade and Pauli-blockade magnetometry. 
(a) Coulomb-blockaded single-level quantum dot
between source (S) and drain (D).
The spin states are split by an offset magnetic field
$B_0$ setting the working point. 
(b,c) Current is flowing through the $\downarrow$ state, 
therefore the current changes by $\delta I$
if the energy of $\downarrow$ is
changed by a small variation $\delta B$ of the magnetic
field.
This allows for measuring $\delta B$ via measuring the current.
The shaded areas in (b) represent the thermally broadened
electronic Fermi-Dirac distributions in the contacts. 
(d) Pauli-blockaded double quantum dot. 
The two ellipsoids represent the $g$-tensors on the two dots.
The $g$-tensors are anisotropic, and have
misaligned principal axes in the two dots, enclosing the misalignment
angle $2\alpha$. 
The working point of the magnetometer is set by the field
$\vec B_0 = (B_{0x},0,B_{0z})$,
and $\delta B_z$ is measured via measuring
the change $\delta I$ [see (e)] 
in the current flowing through the dots. 
(e) Current as a function of $B_z$. 
Black square ($\blacksquare$) denotes the point of $T_0$ blockade.
(f) Two-electron level diagram in the vicinity of $T_0$ blockade. 
Solid and dotted black arrows represent 
magnetic-field-induced coupling matrix elements.
At the $T_0$ blockade ($\blacksquare$), the $T_0$-$S$ coupling
(dotted black arrow) vanishes. 
\label{setup}}
%\vspace{-0.4cm}
\end{figure*}

In this work, we propose and theoretically 
explore two quantum-dot-based 
low-temperature approaches to magnetic-field 
sensing, offering a combination of 
sub-$\mu\textrm{T}/\sqrt{\textrm{Hz}}$ 
magnetic-field sensitivity,  
nanoscale spatial resolution, and conceptual simplicity via all-electrical 
and all-dc operation (i.e., optical or high-frequency electronic
elements are not required).
In both devices,  the magnetic field is measured by measuring
the electric current through the dots.
The first, simpler device we study is
a single quantum dot in the Coulomb blockade regime;
the second one is a 
double quantum dot (DQD) in the Pauli blockade 
regime\cite{Ono-spinblockade}. 
Our main goal is to determine the fundamental limits on the 
achievable magnetic-field sensitivity of these sensors by 
considering the effects of shot noise and the thermal broadening
of the electron distributions in the contacts.
We also discuss the effect of electric potential fluctuations
on the sensitivity,
and highlight the advantages of carbon nanotubes (CNTs) for
realizing the proposed magnetometry principle.

\section{Sensitivity of a current-based magnetometer}
Before presenting the concrete magnetometry schemes, 
we first characterize the shot-noise-limited sensitivity of a
generic magnetometer that is based on the magnetic-field dependence
$I(\vec B)$
of the steady-state 
electric current flowing through a mesoscopic conductor.
We wish to exploit this dependence for measuring small deviations
$\delta \vec B$
of the magnetic field from a pre-set 
`offset field' or `working point' $\vec B_0$.  
We focus on the generic situation 
 when the current is linearly sensitive
to a single Cartesian component of $\delta \vec B$, say,
$\delta B_z$; that is,
$I' (\vec B_0)  \equiv
 \left. \frac{\partial I}{\partial B_z}\right|_{\vec B_0}  > 0$,
 and 
 $\left. \frac{\partial I}{\partial B_x}\right|_{\vec B_0} = 
\left. \frac{\partial I}{\partial B_y}\right|_{\vec B_0} = 0$.
This is always fulfilled if we 
align the $z$ axis with the gradient vector of $I(\vec B)$ at $\vec B_0$.
In this case, the device can be operated as a linear detector of
$\delta B_z$.
In analogy with the sensitivity formula for 
a current-based electrometer\cite{Korotkov-electrometer}, 
we claim  
that the magnetic-field sensitivity of the magnetometer is
characterized by the quantity
\bean
\label{eq:genericsensitivity}
S(\vec B_0) = \frac{\sqrt{e F(\vec B_0) I(\vec B_0)}}{I'(\vec B_0)},
\eean
where $e$ is the absolute value of the 
electron's charge, and 
$F$ is the Fano factor\cite{Blanter2001}, defined as
the ratio of the shot noise and the current.
For more details, see
Appendix \ref{app:sensitivity}. In what follows, we will refer 
to the shot-noise-limited sensitivity $S$ as the sensitivity.

The dimension of $S$ is ${\rm T}/\sqrt{\rm Hz}$. 
A smaller value of 
$S$ implies the ability of resolving smaller differences in $\delta B_z$
in a given measurement time window; that is, an improved performance. 
Equation \eqref{eq:genericsensitivity} is in line with the 
expectation that the sensitivity is improved if the noise is suppressed
($F$ is decreased)
or if the dependence of the current on the magnetic field is
enhanced ($I'(\vec B_0)$ is increased).

\section{Coulomb-blockade magnetometry}
\label{sec:cbm}

Here, we describe and characterize a principle of magnetometry
based on Zeeman-splitting-induced 
changes in the current flowing through a single dot,
as sketched in Fig.~\ref{setup}a,b,c.
The scheme is analogous to the electrometer described in 
Refs.~\onlinecite{Korotkov-electrometer,Korotkov-noise}.

The dot, shown schematically
in Fig.~\ref{setup}a, is gate-voltage-tuned 
to the vicinity of a Coulomb peak,
and a finite magnetic field $\vec B_0 = (0,0,B_0)$
creates a large Zeeman splitting $g \mu_B B_0$,
where $g$ is the material-dependent effective 
$g$-factor of the electron. 
The Zeeman splitting separates the singly occupied $\uparrow$
(spin-up) excited state
from the $\downarrow$ (spin-down) ground state,
as shown in Fig.~\ref{setup}b. 
The magnetic field also creates a Zeeman splitting in the leads. We assume that under a finite source-drain bias voltage, the electron distributions in the leads are thermal
(represented by the shaded regions in Fig.~\ref{setup}b) and are characterized by spin-independent local chemical potentials
(represented as the dashed horizontal lines in Fig.~\ref{setup}b).
A plunger gate voltage and a small source-drain bias voltage 
can be used to tune the energy levels such that  $\downarrow$
is in the source-drain bias window as shown in Fig.~\ref{setup}b. 
Then the 
current flows via sequential tunneling through the transport cycle
$0 \to \downarrow \to 0$, where
0 corresponds to an empty dot,
and $I(B_0)$ is set to the slope of the Coulomb-peak,
see the working point in Fig. \ref{setup}c.
Then, a small $\delta B$ increase in the $z$ component
of the magnetic field lowers the energy
of the current-carrying $\downarrow$ state by $g\mu_B \delta B$, 
and therefore 
increases the current flowing through the dot by the
amount $\delta I$ as shown in Fig.~\ref{setup}c;
measuring this increase $\delta I$ 
of the current will reveal $\delta B$.

This measurement scheme is directionally sensitive in the
following sense. 
Assume that the small change in the magnetic field has all
three  
Cartesian components, $\delta \vec{B} = (\delta B_x, \delta B_y,  
\delta B_z)$, and it is much weaker than the offset field, 
$\delta B \ll B_0$. 
Then the Zeeman splitting 
$g\mu_B \sqrt{(B_0 + \delta B_z)^2 + \delta B_x^2+\delta B_y^2}$
is well approximated by 
$g \mu_B \left(
B_0 + \delta B_z + \frac{\delta B^2}{2 B_0}
\right)$,
that is, it is 
mainly determined by the
component $\delta B_z$ along the offset field.

Now we use simple considerations to estimate the parameter
dependence of the sensitivity $S$, and argue that the
temperature has a strong influence on the optimal sensitivity: 
the latter is degraded as $T$ is increased.
For these considerations, we introduce the characteristic
rate $\Gamma$, describing the tunnel coupling of the dot to the
source and drain leads. 
We propose that at a given $T$, (i) the sensitivity is optimized 
if $h \Gamma$ is comparable to $k_B T$, 
and
(ii) the order of magnitude of the 
optimal sensitivity is estimated as 
\bean
\label{eq:cbmoptimalsensitivity}
S_\text{opt} \sim \frac{\sqrt{h k_B T}}{g\mu_B}.
\eean
The reasoning is 
as follows. 
The height of the Coulomb peak is set by 
the lead-dot tunneling rate $\Gamma$,
$I \sim e\Gamma$, whereas 
the Fano factor is $F \sim 1$.
The slope of the Coulomb peak can be set
by thermal broadening or tunnel broadening: 
$I' \sim \frac{e\Gamma}{\max\{h \Gamma,k_B T\}/g \mu_B }$.
Then, Eq. \eqref{eq:genericsensitivity} implies 
$S \sim \frac{\max\{h \Gamma,k_B T\}}{g\mu_B \sqrt{\Gamma}}$.
On the one hand, 
this has the consequence that for slow tunneling, 
$\Gamma < k_B T$,
the sensitivity $S \propto 1/\sqrt{\Gamma}$ 
decreases with increasing $\Gamma$;
on the other hand, for fast tunneling
$\Gamma > k_B T$, 
the sensitivity $S \propto \sqrt{\Gamma}$
increases with increasing $\Gamma$. 
These imply claims (i) and (ii). 
Using the estimate in Eq.~\eqref{eq:cbmoptimalsensitivity} and
the  values
$g=30$, achievable in clean CNT
dots\cite{Ajiki,Minot,Kuemmeth,Laird-rmp-review},
and $T = 50 \, \text{mK}$, 
we find 
$S_\text{opt} \sim 77\, \text{nT}/\sqrt{\text{Hz}}$.

Now we go beyond the previous estimate and quantify the 
magnetic-field sensitivity of the Coulomb-blockade magnetometry via a simple model. 
The single-electron Hamiltonian of the quantum dot involves the 
on-site energy $\epsilon$ and the Zeeman term: 
$
H_\text{CBM} = \epsilon + \frac 1 2 g \mu_B B \sigma_z,
$
where
$g$ is the effective g factor, $\mu_B$ is the Bohr magneton, and
$B$ is the magnetic field. 
The electronic Fermi-Dirac distributions in the source
and drain leads are characterized by their common 
temperature $T$ and symmetrically biased chemical potentials
$\mu_L = -\mu_R = e V_{sd}/2$ with $V_{sd}$ being the source-drain
bias voltage. 
Lead-dot tunneling rates are set by the rate
$\Gamma$, the level positions, and the Fermi-Dirac distributions
of the leads. 
We describe the transport process by a classical master equation,
neglecting double occupancy of the dot.
The current $I$ and the Fano factor $F$ is evaluated using 
the counting-field method\cite{Bagrets,PhysRevB.71.161301};
details can be found in Appendix \ref{app:countingfield}.
From these, the magnetic-field sensitivity $S$ is calculated from
Eq.~\eqref{eq:genericsensitivity}.

\begin{figure}
\begin{center}
\includegraphics[width=1\columnwidth]{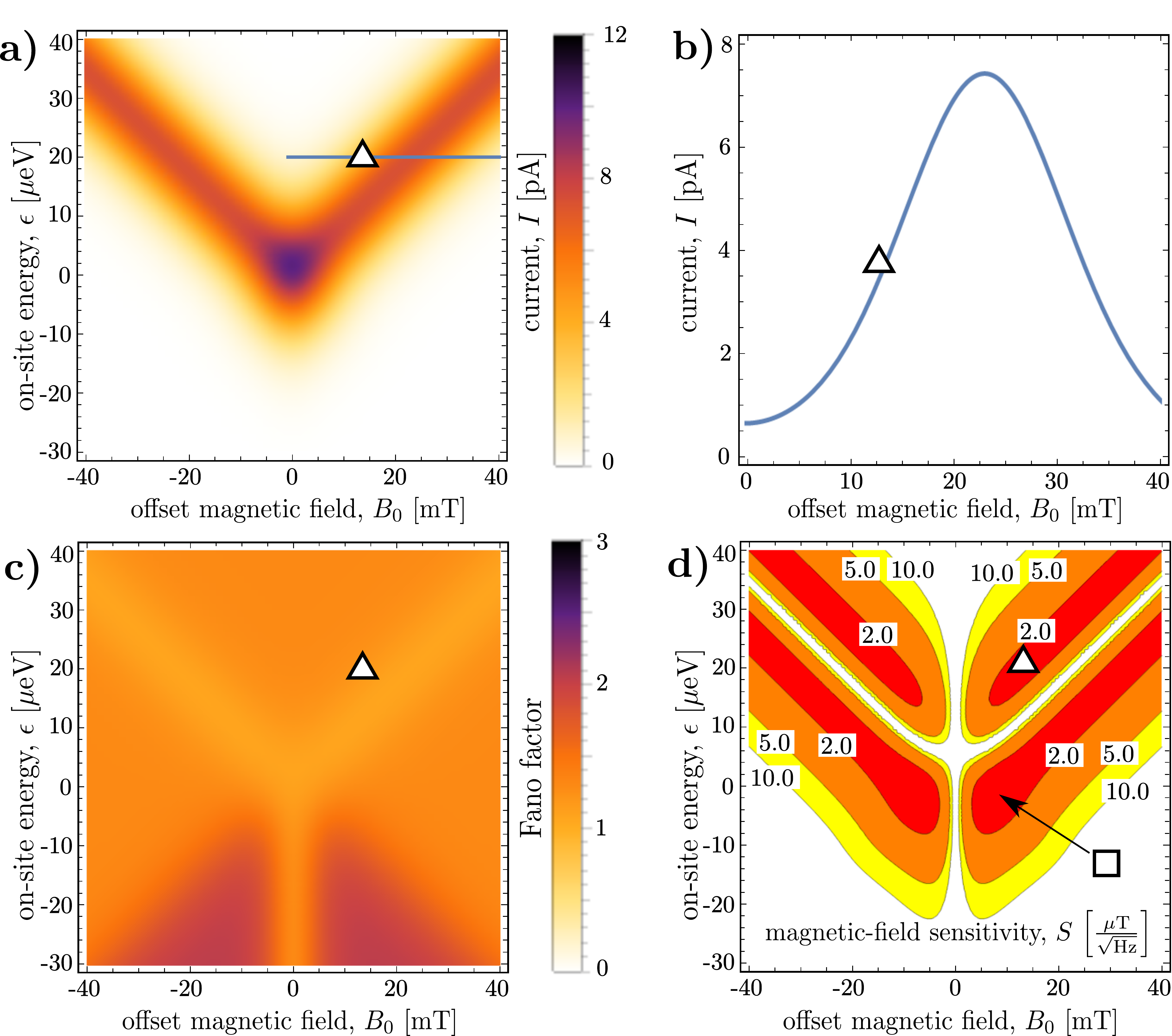}
\end{center}
%\vspace{-0.4cm}
\caption{
Coulomb-blockade magnetometry: Magnetic-field sensitivity of a single 
quantum dot. 
(a,b) Current, (c) Fano factor, and 
(d) sensitivity of a Coulomb-blockade
magnetometer.
(b) Magnetic-field dependence of the current 
along the blue line ($\epsilon=20 \, \mu \text{eV}$) in (a).
The white triangle $\triangle$ denotes a local  minimum of the
sensitivity $S$ along the blue line in (a). The optimal value of sensitivity denoted by the white square $\square$ is $S = 1.4 \ \mu \text{T}/\sqrt{\text{Hz}}$.
Parameters: 
$V_{sd}=8.6\, \mu \text{V}$,
$k_B T=4.3\, \mu\text{eV}$ ($T \approx 50$ mK),
$h\Gamma= 0.86\, \mu \text{eV}$ ($\Gamma \approx 200$ MHz, 
$e\Gamma \approx 33 \, \text{pA}$),
$g = 30$.
}
\label{fig:singleQD_result}
%\vspace{-0.4cm}
\end{figure}

Figures \ref{fig:singleQD_result}a,c,d shows the calculated current
$I$, 
Fano factor $F$ and magnetic-field sensitivity $S$,  respectively, 
as functions of the offset  field $B_0$ and the 
on-site energy $\epsilon$ of the dot, for a given
parameter set (see caption) 
where the thermal energy scale $k_B T$
dominates the tunneling energy scale $h \Gamma$.
Figure \ref{fig:singleQD_result}b shows a horizontal cut 
of the current, $I(B_0, \epsilon=20 \, \text{meV})$, 
along the blue horizontal line in Fig.~\ref{fig:singleQD_result}a.
The dark spot in the centre of Fig.~\ref{fig:singleQD_result}a, 
at $\epsilon \approx 0$ and $B_0 \approx 0$, is a finite-bias
Coulomb peak,
where the two spin levels $\downarrow$ and $\uparrow$ 
are approximately degenerate, both of them is located within
the bias window, and hence both contributes to the current. 
The dark diagonal lines forming the V-shaped region 
in Fig.~\ref{fig:singleQD_result}a 
correspond to Coulomb peaks where
the current is carried by a single spin level,
as shown in Fig~\ref{setup}b: 
a Zeeman splitting $g\mu_B B_0 > k_B T, e V_{sd}$ 
exceeding the 
thermal and voltage broadening is induced between
the spin levels by the offset field $B_0$, 
thereby only the lower-energy spin level contributes to the current.

The typical values of the Fano factor in 
Fig.~\ref{fig:singleQD_result}c
 corroborate our above estimate $F\sim 1$, 
and implies that the Fano factor plays a minor role
in determining the order of magnitude of the magnetic-field sensitivity. 
The sensitivity in 
Fig.~\ref{fig:singleQD_result}d is therefore following
a pattern which can essentially be deduced from 
the pattern of the current, Fig.~\ref{fig:singleQD_result}a:
the sensitivity is good, that is, $S$ is small,
wherever the
derivative of the current with respect to the offset field 
is appreciable;
that is, along the two slopes of  the V-shaped
high-current regions of Fig.~\ref{fig:singleQD_result}a.
In Fig.~\ref{fig:singleQD_result}d, the sensitivity has a local minimum 
$S = 1.7\, \mu\text{T}/\sqrt{\text{Hz}}$ at the white triangle. 

Note that in Fig.~\ref{fig:singleQD_result}d,
the global minimum 
of the magnetic-field sensitivity is at the working point 
marked by the white square ($\square$),
which corresponds to 
$(\epsilon,B_0) \approx (0 \ \mu \text{eV},10 \ \text{mT}) $.
At $\square$, the value of the sensitivity is
$S = 1.4 \ \mu \text{T}/\sqrt{\text{Hz}}$.
We emphasize that the operation of the magnetometer at this 
working point $\square$ is not following the scheme  
that we outlined above and visualized in Fig.~\ref{setup}b.
The scheme of operation in the working point $\square$ 
is discussed in detail in 
Appendix \ref{app:globalminimum}.
In Appendix \ref{app:globalminimum}, 
we also argue that by choosing an appropriate working
point in the vicinity of $\square$, the sensitivity-degrading effects
of electric potential fluctuations can be mitigated. 

To conclude, we argued that the fundamental limit
on the magnetic-field sensitivity of a Coulomb-blockade 
magnetometer is set by the temperature, evaluated and analyzed 
the magnetic-field sensitivity of such a device using a simple model, 
and estimated that 
at experimentally available low temperatures, and
with a large g-factor, e.g., offered by CNT dots,
the optimal magnetic-field sensitivity can 
be below $\mu \text{T}/\sqrt{\text{Hz}}$.

\section{Pauli-blockade magnetometry}

\label{sec:pbm}

Here, we describe an alternative magnetometer
based on a DQD operated in the Pauli blockade 
regime\cite{Ono-spinblockade}.
In such DQDs, the magnetic-field dependence of the current
can be caused by various mechanisms, e.g., 
hyperfine interaction\cite{Koppens-spinblockade,Jouravlev} 
or spin-orbit interaction\cite{Pfund,Danon-spinblockade}.
Here, we focus on one particular mechanism,
where the magnetic-field dependence of the current
is governed by the different and strongly anisotropic g-tensors
in the two dots, as depicted in Fig.~\ref{setup}d.
We will show that if we have 
this feature in the device, then the magnetic-field sensitivity
is optimized in the close vicinity of a special setting that
we call the `$T_0$ blockade'  (see Fig. \ref{setup}e,f, and
below).
Based on a comparison of a recent experiment\cite{FeiPei}
and our corresponding theoretical 
results\cite{PhysRevB.91.045431}, we argue that
DQDs in bent CNTs provide an opportunity 
to meet these requirements.

In the Pauli-blockade regime, a large dc source-drain 
voltage is applied to a serially coupled DQD, and
a dc current might flow via the transport cycle 
$(0,1)\rightarrow(1,1)\rightarrow(0,2)\rightarrow(0,1)$,
where ($N_L$, $N_R$) denotes the number of electrons in the left and 
right dots.
The $(1,1) \rightarrow (0,2)$ transition is blocked 
due to Pauli's exclusion principle if
a (1,1) triplet state becomes occupied during the 
transport process, leading to a complete suppression of the current.
We describe the DQD
by the two-electron Hamiltonian\cite{Jouravlev} 
$H=H_B+H_\textrm{tun}+H_\Delta$.
The interaction of the external homogeneous magnetic field
and the electron spins is 
\begin{equation}
H_B=\frac 1 2 \mu_B \vec B \cdot \left(
	{\mathbf{\hat g}}_L \boldsymbol\sigma_L+
		{\mathbf{\hat g}}_R \boldsymbol\sigma_R
\right).
\end{equation}
Here,
$\boldsymbol\sigma_{L/R}$ is the vector of Pauli matrices 
representing the spins of the electrons.
The $g$-tensors $\mathbf{\hat g}_L$ and $\mathbf{\hat g}_R$
are assumed to have the same principal values  
$g_\perp$, $g_\perp$, $g_\parallel$
in the dots $L$ and $R$.
(This is not a strict requirement, as we discuss in 
section \ref{sec:differentgfactors}.)
Furthermore, the 
principal axes of the g-tensors enclose a small angle $2\alpha \ll 1$,
as shown in Fig.~\ref{setup}d.
Choosing the coordinate system as depicted in 
Fig.~\ref{setup}d, 
the $g$-tensors are given as
$
\mathbf{\hat g}_D =
g_\parallel \vec t_D \circ \vec t_D
+
g_\perp \left( 1- \vec t_D \circ \vec t_D\right),
$
where 
$\vec t_D = (D \sin \alpha ,0,\cos \alpha)$ is
the unit vector pointing along the local principal axis
of $g_\parallel$ in dot 
$D \in (L,R) \equiv (-1,1)$.
Spin-conserving  tunneling between the 
dots is represented by 
$H_t=\sqrt{2} t (\ket{S_g}\bra{S} + \ket{S}\bra{S_g})$, 
where $S$ [$S_g$] is the singlet state in the 
(1,1) [(0,2)] charge configuration. 
The last term $H_\Delta=-\Delta\ket{S_g}\bra{S_g}$
describes  the energy  detuning  between
the  (1,1) and  (0,2)  charge configurations.   
%We set the spin quantization axis along $x$, and
%express $H$ 
%in the basis $T_+$, $T_0$, $T_-$, $S$, $S_g$;
%the corresponding matrix 
%reads \cite{Jouravlev,Danon-organic,PhysRevB.91.045431}
%\newcommand{\BB}{\mathcal{B}}
%\bean
%H = 
%\left(
%\bna{cccc c}
%\BB_s & 0 & 0 & -\BB_{a\perp}/\sqrt{2}& 0  \\
%0 & 0 & 0 & \BB_{a\parallel} & 0 \\
%0 & 0 & -\BB_s & \BB_{a\perp}/\sqrt{2}& 0  \\
%-\BB_{a\perp}/\sqrt{2} & \BB_{a\parallel} & \BB_{a\perp}/\sqrt{2} & 0& \sqrt{2} t \\
%0 & 0 & 0 & \sqrt{2} t & -\Delta 
%\eda
%\right).
%\eean
%where 
%$\mathcal B_s = \frac{1}{2} \mu_B \left[ (\hat{\boldsymbol g}_L + \hat{\boldsymbol g}_R) \vec B\right]_x $
%and 
%$\mathcal B_{a\perp} = \frac 1 2 \mu_B [(\hat{\boldsymbol g}_L - 
%\hat{\boldsymbol g}_R)\vec B]_z$.
Finally, the incoherent tunneling processes from the source
electrode to dot $L$ (from dot $R$ to the drain electrode) are
characterized by the rate $\Gamma_L$ ($\Gamma_R$).

%The transport process is modelled by a classical rate equation,
%where the states are the energy eigenstates of $H$,
%and the rates characterizing the transitions between the states
%are describing single-electron tunneling events between 
%the electrodes and the DQD. 
%For details, see Appendix \ref{app:countingfield}.

In this model, the $T_0$ blockade appears in the case when the magnetic field is aligned with the 
$x$ axis, $\vec B = (B,0,0)$.
Then, taking the spin quantization axis along $x$, which coincides
with the direction of $\vec B$ as well as with that
of the average effective magnetic field
$\frac 1 2 \left(
\mathbf{\hat g}_L + \mathbf{\hat g}_R\right) \vec B$,
the (1,1) triplet state $\ket{T_0} = \frac 1 {\sqrt{2}} \left(
\ket{\uparrow \downarrow} + \ket{\downarrow \uparrow}\right)$
is an energy eigenstate, implying that it is
decoupled from the (0,2) charge configuration
and therefore blocks the current \cite{Jouravlev}. 
This setting is indicated in Fig.~\ref{setup}e and f by the
black square ($\blacksquare$).   
The other four two-electron energy eigenstates all contain
a finite (0,2) component and therefore can decay to 
the single-electron states by emitting an electron to the drain;
hence we call this special case the $T_0$ blockade.
Note that the appearance of the $T_0$ blockade does not
require the equality of the g-tensor principal values in the
two dots, see section \ref{sec:differentgfactors}.

Importantly, the $T_0$ blockade is maintained as long as
the  magnetic field lies in the $xy$ plane. 
It is however lifted by a finite $B_z$ component, 
as the latter couples $T_0$ to the singlet states \cite{Jouravlev}.
Therefore, in the vicinity of the $T_0$ blockade, the current
is mostly determined by $B_z$.
The decay rate of $\bar{T}_0$, that is,
the energy eigenstate that evolves from
$T_0$ as $B_z$ is turned on, is 
given by $\bar\Gamma = 
2\Gamma_R \left| \bra{S_g} \bar{T}_0 \rangle \right|^2$.
We express $\bar{T}_0$ using first-order perturbation theory
in $\mu_B g_\parallel B_z$,
see Appendix \ref{sec:bartzero}.
After a leading-order
expansion in the small angle $\alpha$, and
assuming $g_\perp \ll g_\parallel$, 
fulfilled by the 
realistic parameter set
used below, 
we find 
\bean
\label{eq:overlap}
\left| \bra{S_g} \bar{T}_0 \rangle \right|^2
= 
\left(
 \frac{\mu_B g_\parallel B_z  }{\sqrt 2 t }
\frac{g_\parallel}{g_\perp} \alpha 
\right)^2.
\eean
Note that this result is independent of the 
(1,1)-(0,2) energy detuning $\Delta$;
the consequences of this are discussed in 
section \ref{sec:noise}. 

Assuming that $\bar{\Gamma}$ is the smallest tunnel 
rate in the transport process, 
and making use of the formalism 
of Refs.~\onlinecite{Bagrets,PhysRevB.71.161301},  
we can express the current as
\bean
\label{eq:currentanalytical}
I = 4 e \bar{\Gamma} 
=
8 e \Gamma_R 
\left(
 \frac{\mu_B g_\parallel B_z  }{\sqrt 2 t }
\frac{g_\parallel}{g_\perp} \alpha 
\right)^2,
\eean
and the Fano factor as 
\bean
\label{eq:fanoanalytical}
F=7.
\eean 
Details can be found in Appendix 
\ref{app:countingfield}.

The dependence $I(B_z)$ can be utilized 
for current-based magnetometry. 
Take an offset field with a small 
$z$ component, $\vec B_0 = (B_{0x},0,B_{0z})$.
Consider the task that a small change $\delta B_z$ in
the $z$ component of the field should
be detected via measuring the current.
The corresponding shot-noise-limited sensitivity for the device operated
in the vicinity of the $T_0$ blockade can be 
expressed from Eqs. \eqref{eq:genericsensitivity},
\eqref{eq:currentanalytical} and \eqref{eq:fanoanalytical}:
\bnen 
\label{sensitivitydqd2}
S=\frac{\sqrt{7}}{4}\frac{tg_\perp}{\alpha\mu_Bg_\parallel^2}\frac{1}{\sqrt{\Gamma_\textrm{R}}}.
\eden
For the parameter set $g_\parallel=30$, $g_\perp=1$, 
$\Gamma_R=1$ GHz, $t=5\;\mu$eV, $\alpha=3^\circ$,
realistic for a clean CNT DQD \cite{FeiPei,Laird,PhysRevB.91.045431},
Eq.~\eqref{sensitivitydqd2} implies 
a sensitivity of 
$S\approx 37\frac{\textrm{nT}}{\sqrt{\textrm{Hz}}}$.

\begin{figure}
\begin{center}
\includegraphics[width=1\columnwidth]{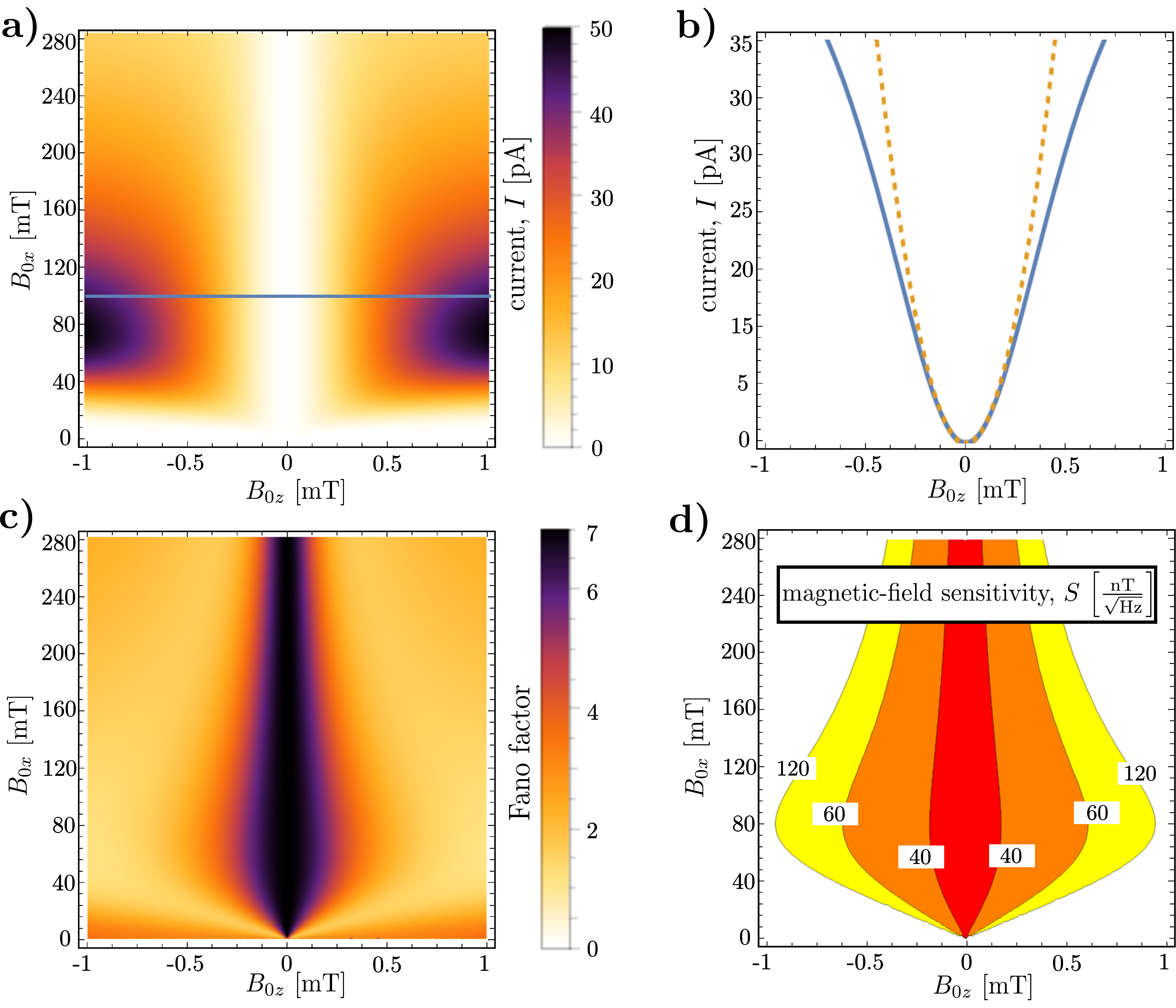}
\end{center}
%\vspace{-0.4cm}
\caption{(Color online)
Pauli-blockade magnetometry:
Magnetic-field sensitivity of a double quantum dot.
Numerically calculated (a) current, (c) Fano factor, and 
(d) magnetic-field sensitivity. 
The suppressed current around $B_{0z} = 0$ in (a) is due to 
$T_0$ blockade. 
(b) Blue line shows the horizontal cut along the blue line 
($B_{0x} = 100 \ \text{mT}$)
in (a).
Dashed orange line shows the corresponding 
perturbative result
Eq.~\eqref{eq:currentanalytical}.
The optimal value of the sensitivity in (d) is 
$S_\text{opt} = 37 \ 
\text{nT}/\sqrt{\text{Hz}}$,
at the point $(B_{0z},B_{0x}) =(0,25 )\ \text{mT}$.
Parameters: 
$g_\parallel=30$, $g_\perp=1$, 
$\Gamma_R=\Gamma_L = 1$ GHz, $t=5\;\mu$eV, 
$\Delta = 0\, \mu\text{eV}$, $\alpha=3^\circ$.
}
\label{fig:pbm}
%\vspace{-0.4cm}
\end{figure}

To explore the achievable magnetic-field sensitivity of the device
in a broader magnetic-field range, and to corroborate our perturbative
results discussed so far, 
we numerically calculated the 
current, the Fano factor, and the sensitivity.
The dependence of these quantities 
on the offset magnetic field are shown 
in Fig.~\ref{fig:pbm}a,c and d, respectively,
for the parameter values given in the figure caption. 
In Fig.~\ref{fig:pbm}b, the blue solid line
shows the current 
along the horizontal cut in Fig.~\ref{fig:pbm}a, whereas
the orange dashed line shows the corresponding perturbative result
Eq.~\eqref{eq:currentanalytical}.
The numerical results in Fig.~\ref{fig:pbm}
do confirm our analytical results 
obtained for the vicinity of the $T_0$ blockade, 
$B_{0z} \approx 0$:
On the one hand, the two curves coincide in that region
in Fig.~\ref{fig:pbm}b; on the other hand, 
the maximum of the Fano factor in Fig.~\ref{fig:pbm}c is
indeed $F=7$ as predicted by Eq.~\eqref{eq:fanoanalytical}.

The key features of the results in Fig.~\ref{fig:pbm} 
are the following.
(i)
Figure \ref{fig:pbm}a shows that the current is
blocked at $B_{0z} = 0$; that is the $T_0$ blockade. 
In its vicinity, the current shows a quadratic dependence of 
$B_{0z}$, as testified by the line cut in Fig.~\ref{fig:pbm}b
and our perturbative result Eq.~\eqref{eq:currentanalytical},
and essentially no dependence on $B_{0x}$.
This implies that indeed, the device as a magnetometer
is directionally sensitive, that is, variations of the current
are caused by variations of the $z$ component of the 
external magnetic field only. 
(ii) 
Fig.~\ref{fig:pbm}d shows that the best sensitivities
are achieved in the vicinity of the central vertical region (black),
which indeed roughly coincides with the $T_0$ blockade
region,
i.e., the white vertical region with suppressed current
in Fig.~\ref{fig:pbm}a and 
the dark $F\approx 7$ vertical region in the Fano factor plot
Fig.~\ref{fig:pbm}c.
(iii) The current in Fig.~\ref{fig:pbm} shows maxima
around $(B_{0z},B_{0x}) = (\pm1 \, \text{mT}, 80\,\text{mT})$. 
The reason for this is that singlet-triplet mixing is most efficient
in these magnetic-field ranges. 
At these dark spots, the local effective magnetic fields 
$\hat{g}_L \vec B$ and $\hat{g}_R \vec B$ in 
the two dots differ significantly, and their energy scales 
$|\mu_B \hat{g}_L \vec B|$ and $|\mu_B \hat{g}_R \vec B|$ 
are comparable to the interdot tunnel amplitude $t$; 
these conditions ensure efficient singlet-triplet 
mixing\cite{Jouravlev} and a large current as seen in Fig.~\ref{fig:pbm}a. 
If the magnetic field is oriented along the same direction, but 
decreased in magnitude, 
$ |\mu_B \hat{g}_{L,R} \vec B| \ll t$,
then the hybridized singlet states with energies around
$\pm \sqrt{2} t$ are hardly mixed with the triplets around
zero energy, hence the latter ones block transport and the 
current decreases, as seen in Fig.~\ref{fig:pbm}a.

To conclude, we characterized the ultimate sensitivity
of a directionally sensitive magnetometry scheme, where the
sharp, magnetic-field-induced variations 
of the current through a Pauli-blockaded 
DQD are used to measure small changes of the external 
magnetic field.
Using parameter values taken from CNT DQD experiments, 
we estimate that the sensitivity of such a device can 
reach a few tens of $\text{nT}/\sqrt{\text{Hz}}$.

\section{Discussion}

\subsection{The role of the temperature}

Based on Eq.~\eqref{eq:cbmoptimalsensitivity}, 
we argued that
the optimal sensitivity of the Coulomb-blockade 
magnetometer is degraded if the
temperature is increased.
On the other hand, the sensitivity of a Pauli-blockade magnetometer 
is not influenced directly by the temperature,
see Eq.~\eqref{sensitivitydqd2}.
The only requirement on temperature for the Pauli-blockade magnetometry scheme
to work 
is that the orbital level spacing 
of the dots should dominate the thermal energy scale.
One remarkable consequence of this fact is that
downsizing the proposed Pauli-blockade magnetometry scheme, e.g., using an atom-sized
DQD based on small donor 
ensembles\cite{Weber_dqd,Weber_spinblockade}, 
which would provide larger orbital level spacings, 
should allow for a higher temperature of operation. 
Such a miniaturization would provide an added benefit: 
enhanced spatial resolution.

\subsection{Optimal sensitivity and dynamical range
for the Pauli-blockade magnetometer}

The optimal value of the sensitivity in Fig.~\ref{fig:pbm}d 
is
$S_\text{opt} = 37 \ 
\text{nT}/\sqrt{\text{Hz}}$,
at the point $(B_{0z},B_{0x}) =(0,25 )\ \text{mT}$.
However, this is not a suitable working point if we want to 
use the device as a linear detector of $\delta B_z$, 
since the dependence of the current on $\delta B_z$ is
quadratic here, 
see Eq.~\ref{eq:currentanalytical} and Fig.~\ref{fig:pbm}b. 
In other words, even though the sensitivity is optimal 
at this working point,  the dynamical range of 
the device as a linear detector of $\delta B_z$ is zero.
Therefore, in practice, the achievable optimal sensitivity 
is slightly degraded with respect to $S_\text{opt}$,
and depends on the desired dynamical range. 
For example, according to Figs.~\ref{fig:pbm}b,d, 
choosing the working point around
$(B_{0z},B_{0x}) = (0.25,100) \, \text{mT}$, 
results in a dynamical range of a few hundred microteslas
and a sensitivity around $50 \, \text{nT}/\sqrt{\text{Hz}}$.

\subsection{The role of electric potential fluctuations}

\label{sec:noise}

Electric potential fluctuations 
might arise from, e.g., gate voltage noise or
the randomly varying occupation of charge traps near the dots. 
These fluctuations modify the parameters
of the quantum dot system; 
for example, they detune the dots' energy levels
from their pre-defined positions. 
This could degrade the sensitivity of 
the Coulomb-blockade magnetometry scheme
presented in Sec.~\ref{sec:cbm}, 
since the noise-induced random detuning 
of the energy levels competes with the
magnetic energy shift which one wants to measure. 

The detailed quantitative characterization of the role of 
electric potential fluctuations is postponed for future work. 
We note, however, that these fluctuations are qualitatively different
from the shot noise treated in this work, in the sense that 
shot noise imposes an unavoidable, fundamental limitation on the
magnetometer's sensitivity, whereas the role and importance
of the electric potential fluctuations is device dependent,
and can probably be mitigated by technological advancements
resulting in noise suppression. 
Also, we emphasize that Coulomb-blockade electrometry,
using similar setups as proposed here, has been successfully
realized\cite{Yoo,Honig2013} 
in spite of the electric potential fluctuations
present in real devices; this fact further supports the 
feasibility of Coulomb-blockade magnetometry.

Furthermore, our
perturbative result \eqref{eq:overlap} suggests that
 Pauli-blockade magnetometry enjoys some protection against electrical 
potential fluctuations: the overlap \eqref{eq:overlap}
and hence the
decay rate 
$\bar{\Gamma}=
2\Gamma_R \left| \bra{S_g} \bar{T}_0 \rangle \right|^2$ 
of $\bar{T}_0$
are independent of the (1,1)-(0,2) energy detuning $\Delta$.
Therefore, even if a weak noise induces fluctuations
of $\Delta$, the decay rate $\bar{\Gamma}$ and 
hence the characteristics of current flow through the device 
remain unchanged, 
and the sensitivity of the magnetometer will not be 
degraded.
We also note that the alternative Coulomb-blockade magnetometry
scheme, corresponding to the vicinity of the working point
$\square$ in Fig.~\ref{fig:singleQD_result}d, also enjoys
protection against electrical potential fluctuations;
see Appendix \ref{app:globalminimum} for a discussion.

\subsection{Different g factors in the two dots}

\label{sec:differentgfactors}

In section \ref{sec:pbm}, we considered a model of a DQD
where the principal values $g_\parallel$ and 
$g_\perp$ of the $g$-tensors are identical in the two dots.
Within this model, the two key qualitative results we have
emphasized were the following.
(i) If the homogeneous magnetic field $\vec B_0$ is
oriented along $x$, then the
$T_0$ blockade sets in, and the dependence of the current on
the component of the magnetic field that is in the $xz$ plane
and perpendicular to $\vec B_0$ allows the measurement
of the latter. 
(ii) The independence of the decay rate $\bar{\Gamma}$ 
from the detuning $\Delta$ indicates that the 
magnetic-field sensitivity of the Pauli-blockade magnetometry is robust against 
electric potential fluctuations. 
We emphasize that these two results are not  
restricted to the case when $g_\perp$ and $g_\parallel$
are identical in the two dots; 
if the g tensors are more generic, but still allow for 
the $T_0$ blockade to appear, then
both properties (i) and (ii) remain. 
The generic condition for the $T_0$ blockade is that 
the local effective magnetic fields on the two dots should have
the same absolute values:\cite{Jouravlev}
$|{\bf{\hat g}}_L \vec B| = |{\bf{\hat g}}_R \vec B|$.
If such a magnetic-field orientation exists for the given g-tensors, 
and the field is oriented along that direction, 
then a 
a single (1,1) two-electron triplet state
is decoupled from the others and blocks the current. 
If we take the spin quantization axis 
along ${\bf{\hat g}}_L \vec B + {\bf{\hat g}}_L \vec B$,  
then this blocking state is in fact $T_0 = \frac{1}{\sqrt{2}} (\uparrow \downarrow + \downarrow \uparrow)$.

\subsection{Mechanical stability requirement for
the Pauli-blockade magnetometer}

If the particular realization of the Pauli-blockade magnetometry discussed here is
used as a scanning probe, then an important requirement
is the mechanical stability of the setup.
This requirement is related to the fact that in the Pauli-blockade magnetometry setup, 
the measured quantity (current) is sensitive to a magnetic-field component that is perpendicular to the offset field. 
Recall that in the Pauli-blockade magnetometry, a homogeneous external magnetic field (offset field) is applied, which is almost aligned with the x axis of the reference frame; for example, $\vec B_0 = (100,0,0.25)$ mT.
Then, measuring the current provides a measurement of $\delta B_z$, that is, the z component of the unknown part of the magnetic field. 
However, in the presence of weak mechanical noise, influencing either the offset field direction or the DQD orientation, their enclosed
angle will change.
For example, consider the change when, due to some mechanical instability, the offset field suffers an unwanted rotation $\vec B_0 \mapsto \vec B'_0$ around the y axis with a small angle $\beta = 1/50 \approx 1$ degree. 
This change will give rise to an offset-field z component $B'_{0z} \approx B_{0x} \sin \beta \approx 2$ mT, far exceeding the original working-point value $B_{0z} = 0.25$ mT. 
This observation highlights the requirement of a high degree of mechanical stability in potential future devices realizing the proposed magnetometry scheme.

\subsection{Realization}

The key elements of the schemes studied here
have already been realized
in CNT-based quantum dot systems, 
suggesting the feasibility of the outlined principles. 
Note that the spin degree of freedom in the models used
in this work correspond to the combined spin-valley states
in CNTs that form twofold degenerate Kramers doublets 
in the absence of a magnetic field\cite{Laird-rmp-review}.

High-quality single and double dots\cite{Laird-rmp-review,Minot,Kuemmeth,ChurchillPRL,ChurchillNPhys,Steele_dqd,Jespersen,Waissman,Benyamini},
including those in bent CNTs\cite{Lai,FeiPei,Laird,Hels}, have been fabricated.
Longitudinal g-factors up to 
$g_\parallel \approx 50$ have been measured\cite{Kuemmeth}, 
and 
Pauli blockade was observed\cite{ChurchillPRL,ChurchillNPhys,FeiPei,Laird}.
Part of the experimental data in a bent-CNT DQD
(see Fig.~S7 in the Supplementary Information of 
Ref.~\onlinecite{FeiPei})
and its comparison with 
theory
(see Fig.~1c in Ref.~\onlinecite{PhysRevB.91.045431}) suggest that
in that device, the $g$-tensors show the  characteristics
required to realize the scheme proposed here, and
that the $T_0$ blockade 
(referred to as `antiresonance' in Ref.~\onlinecite{PhysRevB.91.045431})
have already been observed.

Scanning-probe devices based on quantum dots\cite{Yoo}, including
a gate-tunable CNT quantum dot\cite{Honig2013}, have 
already been fabricated and used for imaging electric 
fields.
Besides demonstrating that CNT dots are compatible with 
scanning-probe technology, the latter fact also highlights the 
opportunity to use quantum dots as hybrid sensors, 
with in situ switching capability between electric-field and
magnetic-field sensing.

Nuclear spins can provide a strong magnetic noise for 
the electron spins via the hyperfine interaction\cite{Hanson}, 
and therefore their presence is 
expected to degrade the performance of
the magnetometer setups described in this work.
This effect might be detrimental for magnetic-field
sensing with dots in frequenty used III-V semiconductor hosts, 
such as GaAs, InAs or InSb,
where the strength of the hyperfine-induced magnetic
noise is in the millitesla range\cite{Koppens-spinblockade}.
It is natural to expect that this noise strength also sets the lower limit
of the magnetic-field resolution of the magnetometry schemes 
described here.
One strategy to overcome this limitation is to use dynamical 
techniques to control the nuclear-spin ensemble
\cite{Reilly,Shulman}, and
thereby reduce the randomness of the corresponding effective
magnetic field acting on the electrons. 
Another strategy is to use a host material with 
a low abundance of nuclear spins. 
For example, natural CNTs 
offer a low, 1\% abundance of nuclear spins,
which can be further lowered using isotopic purification, 
mitigating the harmful effect of the nuclear-spin bath. 
We also note that isotopic purification of 
carbon (diamond) and silicon 
has already been proven\cite{Itoh} to be
a successful way to substantially reduce hyperfine-induced
magnetic noise and thereby effectively prolong electron-spin
coherence times in these materials.

\section{Conclusions} 

We have presented two magnetic-field sensing
principles, which are based on electron transport in 
Coulomb- and Pauli-blockaded quantum dots. 
We demonstrated that the fundamental
limit of the magnetic-field sensitivity in such devices
can be below $ \mu \textrm{T}/\sqrt{\textrm{Hz}}$.
A carbon nanotube might be optimal as a host material.
Downsizing the proposed scheme,
e.g., using an atomic-sized double quantum dot based on small
donor ensembles, might lead to 
further improved spatial resolution,
and, as a consequence of higher orbital level spacings, 
a broader temperature range of operation.

\begin{acknowledgments}
We thank K. Nowack and Sz. Csonka for useful discussions.
We acknowledge funding from 
the EU Marie Curie Career Integration Grant CIG-293834 (CarbonQubits),
the OTKA Grant PD 100373, 
the OTKA Grant 108676, and the EU ERC Starting Grant 
CooPairEnt 258789.
A.~ P.~ is supported by the 
J\'anos Bolyai Scholarship of the Hungarian Academy of Sciences.
\end{acknowledgments}

\appendix

\section{Derivation of the sensitivity
formula}

\label{app:sensitivity}

Here we show that the shot-noise-limited magnetic-field sensitivity 
of a current-based magnetometer is characterized by 
Eq.~\eqref{eq:genericsensitivity}.
The notation used here follows that introduced in the main text.

Charge flow through the device has a random character, 
since electron tunneling between the leads and the conductor
occur in a random fashion.
To describe this randomness, we use $P(N,\tau)$,
the probability that in the time window between
$t=0$ and $t=\tau$, the number of
electrons entering the drain contact is $N$.
The normalization condition is that $\sum_{N\in \mathbb{Z}} P(N,\tau) = 1$ 
for any $\tau$. 
Henceforth we assume that the device is in its steady state, 
and that $P(N,\tau)$  corresponds to this steady state.

We assume that the current is measured by counting 
the  electrons entering the drain during the 
measurement time window $t\in [0,\tau]$. 
The number of transmitted electrons is a random variable;
let us characterize its probability distribution by 
its average $\mathcal N(\tau) = \sum_{N\in \mathbb{Z}} N P(N,\tau)$ and
its variance 
$\sigma^2(\tau) = \sum_{N\in\mathbb{Z}} \left[ N - \mathcal N(\tau)\right]^2 P(N,\tau)$.
Note that in the steady state, both quantities are proportional to 
the measurement time $\tau$ 
(see Refs.~\onlinecite{Bagrets,PhysRevB.71.161301} 
and Eqs. \eqref{atlag} and \eqref{szoras} below), 
which motivates
the definitions of the electric current $I$ and the 
Fano factor $F$,
\bean
\label{eq:currentdef}
I &=& -e \mathcal N(\tau) / \tau,\\
\label{eq:fanodef}
F &=& \sigma^2(\tau) / \mathcal N(\tau),
\eean
which are $\tau$-independent quantities.
Here, $e$ denotes the absolute value of the electron charge.

In the considered generic magnetometer, both 
the average and the variance depend on the magnetic field,
$\mathcal N(\tau) = \mathcal N(\tau,B)$ 
and 
$\sigma^2(\tau) = \sigma^2(\tau,B)$. 
The task is the following.
Assume that $I(B)$ is known,
and that a known working-point magnetic field $B_0$ 
is switched on.
There is also an unknown small component $\delta B$
of the magnetic field $B=B_0 + \delta B$.
We wish to determine $\delta B$ by measuring 
the number of  electrons entering the drain contact in a
given measurement time window $\tau$. 
That is possible, if the  change of the
average number of transmitted
electrons induced by $\delta B$ exceeds the standard deviation
of the transmitted electrons at the working point.
This condition is formalized as
\bean
\label{eq:sensitivitystartingpoint}
|\mathcal N(\tau,B_0+\delta B) -\mathcal N(\tau,B_0) | > 
\sigma(\tau,B_0).
\eean 

This condition leads to a natural definition of the 
magnetic-field sensitivity in accordance with 
Eq.~\eqref{eq:genericsensitivity}:
After
(i) linearizing Eq.~\eqref{eq:sensitivitystartingpoint}
in $\delta B$,  
(ii) expressing the resulting formula
in terms of the current $I$ and 
Fano factor $F$ defined 
in Eqs.~\eqref{eq:currentdef} and \eqref{eq:fanodef}, and
(iii) rearranging the two sides of the resulting equation,
we find that the minimal change in magnetic field
that can be resolved by the current measurement is constrained by
\bean
\label{eq:deltaBmin}
| \delta B | > \frac{1}{\sqrt{\tau}}
\frac{\sqrt{e F(B_0)  I(B_0)}}{| I'(B_0)|}
%\equiv
%\frac{1}{\sqrt{\tau}} S(B_0),
\eean
where the prime denotes the derivative with respect to the 
magnetic field. 

We conclude that the second fraction on the right hand side of Eq. 
\eqref{eq:deltaBmin}, defined as 
the magnetic-field sensitivity $S$ in Eq.~\eqref{eq:genericsensitivity}, does indeed characterize the performance
of the magnetometer:
for a given measurement time $\tau$, the  minimum resolvable 
change in the magnetic field is $ S/ \sqrt{\tau}$.

\begin{figure}
\begin{center}
\includegraphics[width=1\columnwidth]{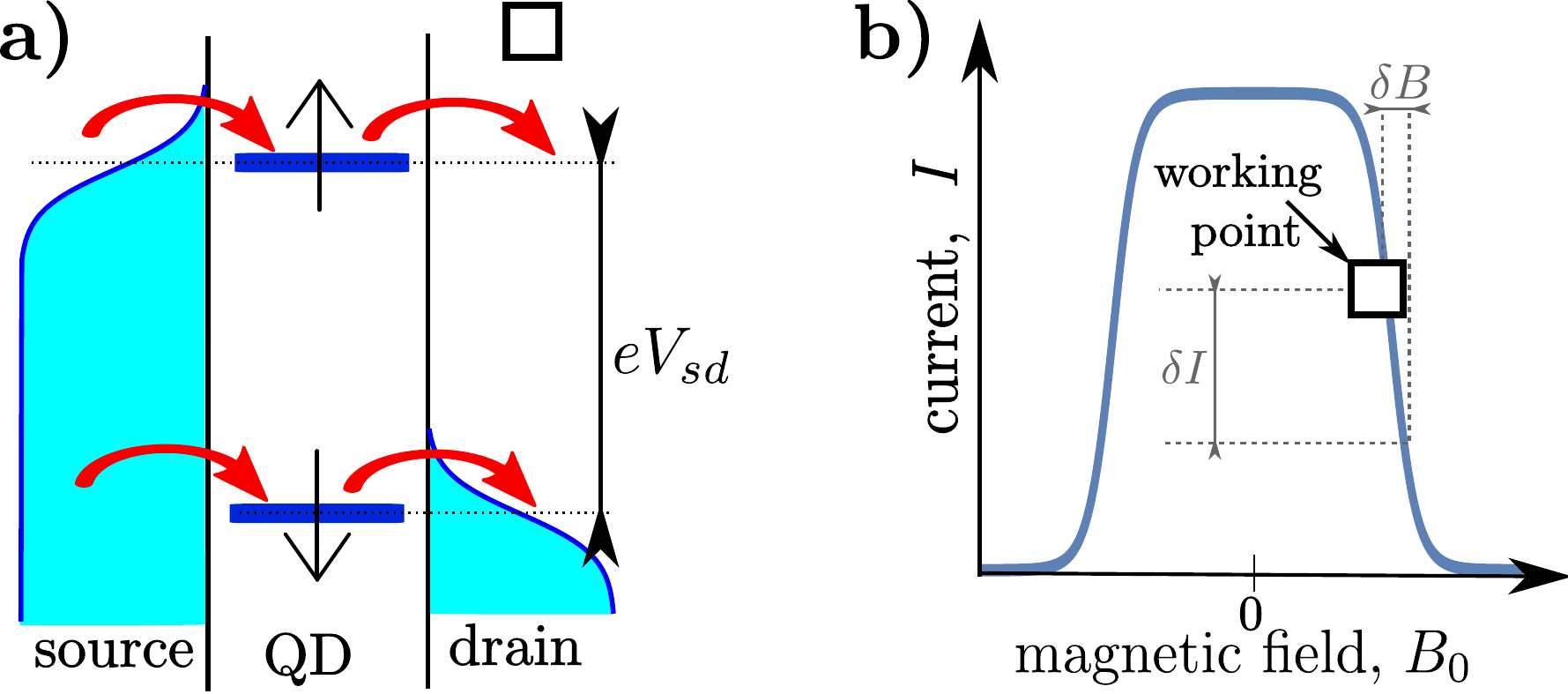}
\end{center}
%\vspace{-0.4cm}
\caption{
Alternative operation of a Coulomb-blockade magnetometer.
(a) Level diagram corresponding to the optimal-sensitivity 
working point
marked by $\square$ in Fig.~\ref{fig:singleQD_result}c.
Here, the offset magnetic field and the source-drain bias
are chosen so that the two spin levels are aligned with 
the chemical potentials of the two leads.
(b) Magnetic-field dependence of the current in the vicinity
of the working point $\square$. 
The width of the current plateau is 
$\sim e V_{sd}/g\mu_B$, that is, it is 
set by the source-drain bias $V_{sd}$. 
The width of the slopes connecting the zero-current 
range with the current plateau is $\sim k_B T / g \mu_B$, 
that is, it is set by the temperature $T$. 
}
\label{fig:cbm2}
%\vspace{-0.4cm}
\end{figure}

\section{An alternative scheme 
of Coulomb-blockade magnetometry}
\label{app:globalminimum}

In Fig.~\ref{fig:singleQD_result}d, 
the global minimum of the magnetic-field 
sensitivity $S$ is at the point marked by the
white square $\square$, that is, at 
$(\epsilon,B_0) \approx (0 \, \mu\text{eV},10 \, \text{mT})$.
As mentioned in section \ref{sec:cbm}, the operation 
of the magnetometer at this working point 
does not follow the scheme
proposed there:
the level alignment is not as shown in 
Fig.~\ref{setup}b.
We show the schematic level alignment for
the working point $\square$ in 
Fig.~\ref{fig:cbm2}a:
each Zeeman-split level 
is located in the vicinity of the chemical potential 
of one of the leads. 
In this situation, the dependence of the current on the magnetic
field is due to the following reason. 
A small change $\delta B>0$ in the magnetic field 
shifts the energy of the $\downarrow$ ($\uparrow$) 
state downwards (upwards),
therefore the outgoing (incoming) rate towards the drain (from the
source) decreases. 
That is, transport through
both levels is suppressed, and therefore the current decreases,
as shown schematically in Fig.~\ref{fig:cbm2}b.

\begin{figure*}
\begin{center}
\includegraphics[width=2\columnwidth]{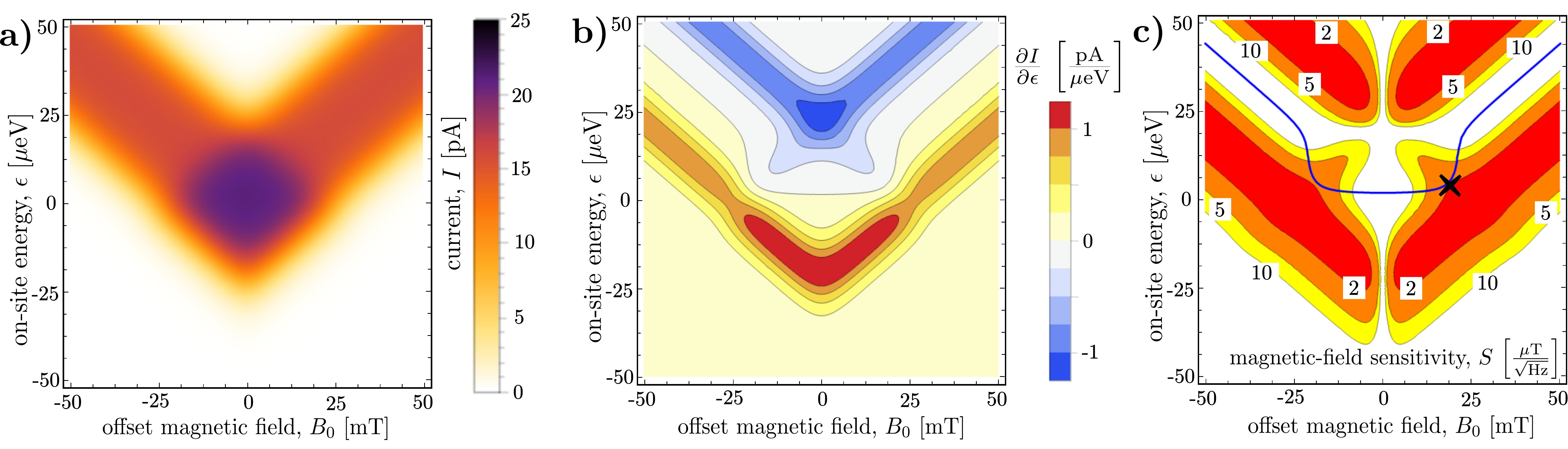}
\end{center}
%\vspace{-0.4cm}
\caption{
Charge-noise-resilient working point of 
the Coulomb-blockade magnetometer.
(a) Current $I(B_0,\epsilon)$ as a function of 
the magnetic field and the quantum dot on-site energy.
(b) 
Derivative of the current with respect to the on-site energy.
Working points along the contour 
$\frac{\partial I}{\partial \epsilon} = 0$ are expected to have
a reduced susceptibility to charge noise (i.e., fluctuations 
of $\epsilon$). 
(c) 
Magnetic-field sensitivity $S(B_0,\epsilon)$.
The solid blue line is the $\frac{\partial I}{\partial \epsilon} = 0$ 
contour.
The black star is the working point where the 
magnetic-field sensitivity is optimized 
($S_\text{opt} \approx 2\, \mu\text{T}/\sqrt{\text{Hz}}$)
along the contour. 
Parameters:
same as in Fig.~\ref{fig:singleQD_result}, 
except that 
$V_{sd}=43\, \mu \text{V}$.
\label{fig:cbm2results}}
%\vspace{-0.4cm}
\end{figure*}

We expect that the effect of electrical potential fluctuations, 
that is, the effect of the noise of the on-site energy $\epsilon$,
can be mitigated by choosing a special working 
point in the vicinity of $\square$. 
We illustrate this opportunity using Fig.~\ref{fig:cbm2results}. 
Figure \ref{fig:cbm2results}a shows the current 
as a function of the 
magnetic field $B_0$ and the on-site energy $\epsilon$,
whereas Fig.~\ref{fig:cbm2results}b shows 
the derivative of the current with respect to the 
on-site energy, $\frac{\partial I}{\partial \epsilon}$.
A strongly reduced effect of $\epsilon$-noise on the magnetic-field 
sensitivity is expected in the working points along the contour
where 
$\frac{\partial I}{\partial \epsilon} = 0$
(see Fig.~\ref{fig:cbm2results}b).
In order to simultaneously 
exploit the opportunity for charge-noise resilience and
obtain a good magnetic-field sensitivity $S$, one should
find the minimum of $S$ along the 
$\frac{\partial I}{\partial \epsilon} = 0$ contour. 
The resulting optimal working point is denoted with the
black star in Fig.~\ref{fig:cbm2results}c, 
where the sensitivity is shown together with the 
$\frac{\partial I}{\partial \epsilon} = 0$ contour
(solid blue line). 
The sensitivity corresponding to this optimal working point
is $S_\text{opt} \approx 2\,  \mu \text{T}/\sqrt{\text{Hz}}$.

\section{Calculating current and shot noise using rate equations}
\label{app:countingfield}

Here, we present the counting-field 
method\cite{Bagrets,PhysRevB.71.161301}
we used to
calculate the current and the shot noise of the 
considered setups, 
and outline the derivation of the specific results 
for the Coulomb-blockade and Pauli-blockade
magnetometers. 

\subsection{Generic framework to calculate the current
and the shot noise: the counting-field method}

The transport process consists of single-electron tunneling events
between the source and the conductor and between the 
conductor and the drain.
Electrons in the conductor can occupy $M$ different states; 
for example, 
in the Coulomb-blockade magnetometer 
setup we consider in section \ref{sec:cbm}
we have $M=3$, as
there are three electronic states involved in the transport cycle:
the empty dot, and the two Zeeman-split single-electron 
states $\downarrow$ and $\uparrow$. 
We assume that the number of charges arriving to the drain
is being measured from some initial point $t=0$ in time. 
Then, the random character of the tunneling events 
is described by the probability density function $P_i(N,\tau)$
where $i=1,\dots,M$; $N \in \mathbb{Z}$; $\tau >0$: 
this is the probability of that at time $\tau$, the conductor 
is in its $i$the electronic state, and $N$ electrons have arrived
to the drain since $t=0$. 
Note that the relation between the probability $P_j(N,\tau)$ introduced
here and the probability $P(N,\tau)$ introduced
in section \ref{app:sensitivity} 
is simply $P(N,\tau) = \sum_{i=1}^M P_i(N,\tau)$.
The normalization condition reads
$\sum_{i=1}^{M} \sum_{N \in \mathbb{Z}} P_i (N,\tau) = 1$
for all $\tau$.
We introduce the vector 
$\mathbf{P}(N,\tau) =\left(P_1(N,\tau), \dots,
P_M(N,\tau)\right)$.
The time evolution of
$\mathbf{P}(N,\tau)$ is governed by a rate equation 
and depends on the initial (in general, mixed) electronic state
characterized by $\mathbf{P}(0,0)$. 
However, here we focus only on the steady-state 
current and shot noise, and
disregard transient effects. 
%Note also that the connection between 
%the  $P_i(N,\tau)$ introduced here and $P(N,\tau)$ used in 
%[SENS] is simply $P(N,\tau) = \sum_{i=1}^{M} P_i(N,\tau)$. 

The effect of random single-electron tunneling events 
is described by the following rate equation:
\bnen
\label{eq:Nresolvedrateequation}
\partial_\tau \mathbf{P}(N,\tau)=\boldsymbol\Gamma_0\mathbf{P}(N,\tau)+\boldsymbol J^+\mathbf{P}(N-1,\tau)+\boldsymbol J^-\mathbf{P}(N+1,\tau),
\eden
where the 
rate matrices $\boldsymbol \Gamma_0$, $\boldsymbol J^+$ and
$\boldsymbol J^-$ have size $M\times M$.
The matrix $\boldsymbol J^+$ ($\boldsymbol J^-$) 
represents single-electron tunneling events from the 
conductor to the drain
(from the drain to the conductor).
For a concrete example for such a rate equation, see
Eq. \eqref{eq:cbmratequation}, the one 
for the Coulomb-blockade magnetometer. 

As the rate matrices are independent of $N$, 
the infinite coupled set \eqref{eq:Nresolvedrateequation}
of differential equations can be partially decoupled 
with the following Fourier transformation:
\bnen\label{fourier}
\tilde{\mathbf{P}}(\chi,\tau)=\sum_{N \in \mathbb{Z}}\mathbf{P}(N,\tau)\;e^{iN\chi}, 
\eden
where $\chi \in ]-\pi,\pi]$.
Acting on Eq. \eqref{eq:Nresolvedrateequation} with
this Fourier transformation yields
\bnen \label{master}
\partial_\tau 
\tilde{\mathbf{P}}(\chi,\tau)=
(\boldsymbol\Gamma_0+
\boldsymbol J^+e^{i\chi}+
\boldsymbol J^-e^{-i\chi})
\tilde{\mathbf{P}}(\chi,\tau)
= \boldsymbol{M} (\chi) \tilde{\mathbf{P}}(\chi,\tau).
\eden
Here the second equality is the definition of $\boldsymbol M(\chi)$,
and
$\chi$ is referred to as the \emph{counting field}. 
For every value of $\chi$, Eq. \eqref{master} represents
a coupled set of $M$ differential equations.

As shown in Refs.~\onlinecite{Bagrets,PhysRevB.71.161301},
the steady-state current and the steady-state shot noise 
are both 
related to the eigenvalue branch $\lambda_1(\chi)$ of
$\boldsymbol M(\chi)$ 
that fulfills $\lambda_1(\chi=0) = 0$.
The average number of transmitted electrons in the steady
state is 
\bnen 
\mathcal N(\tau)=
\sum_{N\in \mathbb{Z}} N P(N,\tau)
\label{atlag}
= -i\left.\tau \frac{\partial \lambda_1(\chi)}{\partial \chi}\right|_{\chi=0},
\eden
whereas the variance is
\bnen \label{szoras}
\sigma^2(\tau)= \sum_{N\in \mathbb{Z}} 
\left[ N^2-\mathcal N^2(\tau) \right] P(N,\tau) =
-\left.\tau \frac{\partial^2 \lambda_1(\chi)}{\partial \chi^2}\right|_{\chi=0}.
\eden
These equations are then used to express 
the current via Eq.~\eqref{eq:currentdef} and 
the Fano factor via Eq.~\eqref{eq:fanodef}. 
To obtain the numerical results
shown in Figs.~\ref{fig:singleQD_result} and \ref{fig:pbm},
we have numerically evaluated the first
and second derivatives
of $\lambda_1(\chi)$ after numerical diagonalization of the
matrix $\boldsymbol M(\chi)$ at $\chi = 0$ 
and $\chi = \pm 10^{-5}$.

%
%  and the 
%The main results of this section
%are Eqs. \eqref{atlag} and \eqref{szoras}.
%They are valid only in the long-time limit 
%$\tau\gg T_{\rm{relax}}$ when the stationary state of the QD is 
%approached. We can establish $T_{\rm{relax}}$ from the 
%$\boldsymbol{M}$ matrix, 
%because all exponential terms in the solution of Eq.(\ref{master}) 
%beside $a(\chi)e^{\lambda_0(\chi)\tau}$ have to vanish. 
%Consequently, $T_{\rm{relax}}\sim\frac{1}{\lambda_1(\chi=0)}$, 
%where  $\lambda_1$ is the second largest eigenvalue of the 
%$\boldsymbol{M}$ matrix.

\subsection{Rate equation for the Coulomb-blockade magnetometer}

The rate equation describing electron transport through
the Coulomb-blockade magnetometer, i.e., a single quantum dot,
reads: 
\bean \label{eq:cbmratequation}
\dot P_\sigma(N,\tau) &=& - \sum_s \Gamma_{\sigma,s}^{\textrm{out}} P_\sigma(N,\tau) + \Gamma_{\sigma,L}^{\textrm{in}} P_0(N,\tau) + \nonumber\\
&+& \Gamma_{\sigma,R}^{\textrm{in}} P_0(N+1,\tau),\nonumber\\
\dot P_0(N,\tau) &=& \sum_{\sigma} \Gamma_{\sigma,R}^{\textrm{out}} P_\sigma(N-1,\tau) + \sum_{\sigma} \Gamma_{\sigma,L}^{\textrm{out}} P_\sigma(N,\tau) -\nonumber\\
&-&\sum_{\sigma,s} \Gamma_{\sigma,s}^{\textrm{in}} P_0(N,\tau),
\eean
where $\sigma \in \{\downarrow=-1,\uparrow=+1\}$ and the indices of the leads are $s \in \{L,R\}$.
Here, 
$P_0(N,\tau)$ is the joint probability of the event that the dot is empty 
at time $\tau$ and $N$ electrons
have entered the drain between $t = 0$ and $t = \tau$. 
Similarly, $P_\uparrow(N,\tau)$ $\left[P_\downarrow(N,\tau)\right]$ is the joint probability of the event that the dot is occupied with a
spin-up [spin-down] electron at time $\tau$, and $N$ electrons
have entered the drain between $t = 0$ and $t = \tau$. 
Furthermore, the transition rates $\Gamma_{\sigma,s}^{\textrm{in}}=\Gamma f_{\textrm{FD}}(\epsilon_\sigma-\mu_s)$ and $\Gamma_{\sigma,s}^{\textrm{out}}=\Gamma \left(1-f_{\textrm{FD}}(\epsilon_\sigma-\mu_s)\right)$ are expressed using the lead-dot tunnelling rate $\Gamma$ and the Fermi-Dirac distribution $f_{\textrm{FD}}(x)=\frac{1}{e^{x/(k_BT)}+1}$. The chemical potentials are biased symmetrically $\mu_L=-\mu_R=\frac{eV_{sd}}{2}$. Due to the Zeeman splitting, the spin-dependent energy of the single occupied dot is $\epsilon_\sigma=\epsilon+\frac{1}{2}\sigma g\mu_BB$. 

Using this rate equation, we  apply the counting-field
method outlined above to calculate the current and the
Fano factor. 
We used analytical results for the 
current to generate Figs.~\ref{fig:singleQD_result}a,b
and \ref{fig:cbm2results}a,  \ref{fig:cbm2results}b. 
The Fano factor shown in \ref{fig:singleQD_result}c,
as well as the sensitivities shown in 
Figs.~\ref{fig:singleQD_result}d and \ref{fig:cbm2results}c, 
are evaluated numerically.

\subsection{Rate equation for the Pauli-blockade magnetometer}

Here, we present the rate equation we use to describe
the transport process through the considered Pauli-blockaded 
double quantum dot. 
The rates of 
$(0,1)\rightarrow(1,1)$ [$(0,2)\rightarrow(0,1)$]
transitions  are parametrized by the rate
$\Gamma_L$ [$\Gamma_R$],
describing single-electron tunneling from the source
to the left dot
[from the right dot to the drain]. 
These considerations result in the following classical master equation \cite{PhysRevB.88.235414}:
\bean \label{master2}
\dot P_j(N,\tau) &=& - 2\Gamma_R v_j  P_j(N,\tau) + \frac 1 2 \Gamma_{\rm L} (1-v_j)  P_6(N,\tau),
\nonumber\\
\dot P_6(N,\tau) &=& - 2 \Gamma_{\rm L} P_6(N,\tau) + 2\Gamma_R \sum_{j=1}^5 v_j  P_j(N-1,\tau).
\eean
Here,
$P_j(N,\tau)$ with $j\in \{1,2,3,4,5\}$ 
is the joint probability of the event that the $j$-th two-electron 
energy eigenstate $\Psi_j$ of the double-dot Hamiltonian 
$H=H_B+H_\textrm{tun}+H_\Delta$ 
(see main text) is occupied at time $\tau$
and $N$ electrons have entered the drain between $t=0$ and
$t= \tau$.
Similarly, $P_6(N,\tau)$ is the joint probability of the event
that a (0,1) state is occupied, 
and $N$ electrons have entered the drain between
$t=0$ and $t=\tau$.
Furthermore, $v_j$ [$1-v_j$] is the weight of the 
two-electron energy eigenstate $\Psi_j$ 
in the (0,2) [(1,1)] charge configuration,
that is, 
$v_j=\left|\langle \Psi_j|S_g\rangle\right|^2$;
note that $\sum_{j=1}^5 v_j=1$.

\subsection{Perturbative analytical results for
the current and the shot noise in the vicinity of $T_0$ blockade}

In the description of the Pauli-blockade magnetometer
in section \ref{sec:pbm},
we quote analytical results
for 
the electrical current, Eq.~\ref{eq:currentanalytical}, 
and the Fano factor, Eq.~\ref{eq:fanoanalytical}.
These results are valid in the case when $B_z$ is sufficiently small,
that is, when the decay rate 
$\bar{\Gamma}$ of the state $\bar{T}_0$ is much smaller than
any other transition rate in the rate equation \eqref{master2}.
To arrive to these analytical results, 
we use the perturbative technique applied in 
Ref.~\onlinecite{PhysRevB.71.161301}.
The key steps are:
(i) The size of $\boldsymbol M(\chi)$ is $6\times 6$, 
hence the
characteristic polynomial of $\boldsymbol M(\chi)$
is of sixth order, $C(\lambda, \bar\Gamma) = \sum_{i=1}^6
c_i(\bar \Gamma) \lambda^i$;
here the coefficients $c_i(\bar \Gamma)$ are 
explicitly determined by the matrix elements of $\boldsymbol M(\chi)$.
(ii) For $\bar \Gamma  = 0$, the state $T_0$ is the steady state, 
with the corresponding eigenvalue 
$\lambda_1(\bar \Gamma =0) = 0$. 
(iii) For small but nonzero $\bar \Gamma$, the eigenvalue $\lambda_1$
developing from $\lambda_1(\bar \Gamma =0) = 0$ is
assumed to be linear in $\bar \Gamma$, that is,
of the form $\lambda_1 = \alpha \bar \Gamma$.  
(iv) Substituting $\lambda = \alpha \bar \Gamma$ to
the characteristic polynomial $C(\lambda,\bar \Gamma)$,
first-order Taylor-expanding the latter in $\bar \Gamma$,
and equating the result to zero,
yields the linear equation 
\bean
C(\lambda,\bar\Gamma) \approx 
\left. \frac{d c_0(\bar \Gamma)}{d \bar \Gamma}
	\right|_{\bar \Gamma=0} \bar \Gamma
+ c_1(0) \alpha \bar \Gamma
=0 .
\eean
Solving this for $\alpha$ yields
\bean
\lambda_1(\chi) = \alpha \bar \Gamma = 
- \frac{\left. \frac{d c_0(\bar \Gamma)}{d \bar \Gamma}
	\right|_{\bar \Gamma=0} }{c_1(0)} \bar \Gamma
	=
	- 4 \bar \Gamma \frac{1-e^{i\chi}}{4-3 e^{i\chi}}.
\eean
This result was used, together with 
Eqs. \eqref{atlag} and \eqref{szoras} to obtain the results 
for the current and the Fano factor.

\section{Resilience to charge noise: detuning-independent
decay rate of $\bar{T}_0$}

\label{sec:bartzero}

Here we provide the derivation of Eq.~\eqref{eq:overlap}
of the main text. 
As stated there, we use first-order perturbation theory to arrive 
to the result. 
Nevertheless, we do present the derivation here 
as it involves one nontrivial technical step:
Usually, the application of perturbation theory is based on the
knowledge of the eigensystem of the unperturbed Hamiltonian;
we do not have that knowledge here, which necessitates
an alternative approach to arrive to the result.

\newcommand{\BB}{\mathcal B}
We start by evaluating the two-electron Hamiltonian 
$H=H_B + H_\textrm{tun}+H_\Delta$
(see main text) 
exactly at the $T_0$ blockade, that is, at 
$\vec B = (B,0,0)$. 
We set the spin quantization axis along $x$, and
express $H$ 
in the basis $T_+$, $T_-$, $S$, $S_g$, $T_0$;
the corresponding matrix 
reads \cite{Jouravlev,Danon-organic,PhysRevB.91.045431}
\bean
H_0 = 
\left(
\bna{cccc | c}
\BB_s & 0 & -\BB_{a\perp}/\sqrt{2}& 0 & 0 \\
0 & -\BB_s & \BB_{a\perp}/\sqrt{2}& 0 & 0 \\
-\BB_{a\perp}/\sqrt{2} & \BB_{a\perp}/\sqrt{2} & 0& \sqrt{2} t & 0 \\
0 & 0 & \sqrt{2} t & -\Delta & 0 \\
\hline
0 & 0 &  0 & 0 & 0 
\eda
\right).
\eean
where 
$\mathcal B_s = \frac{1}{2} \mu_B \left[ (\hat{\boldsymbol g}_L + \hat{\boldsymbol g}_R) \vec B\right]_x $
and 
$\mathcal B_{a\perp} = \frac 1 2 \mu_B [(\hat{\boldsymbol g}_L - 
\hat{\boldsymbol g}_R)\vec B]_z$.
The upper left $4\times 4$ block of $H_0$ will be referred
to as $H'_0$.
In $H_0$, the unpolarized triplet $T_0$ is decoupled from the
other states and therefore blocks the current.
A finite magnetic-field component along $z$,
entering the Hamiltonian as a perturbation
$H_1 = \BB_{a\parallel} (\ket{T_0}\bra{S} + h.c.)$, 
mixes $T_0$ with the other states. 
Here $\BB_{a\parallel} = 
\frac 1 2 \mu_B [(\hat{\boldsymbol g}_L - 
\hat{\boldsymbol g}_R)\vec B]_x$,
and we will denote 
the perturbed $T_0$ state by $\bar{T}_0$.

We were unable to find  
analytical expressions for the eigensystem
of $H'_0$, which would have provided a convenient starting point
for doing perturbation theory in $H_1$. 
However, the overlap $\bra{S_g} \bar{T}_0 \rangle$ we are looking for,
which determines the decay rate of the blocking state $\bar{T}_0$, 
can be expressed in an alternative way.
Start with the standard first-order perturbative
formula for the energy eigenstates:
\bean
\langle S_g \ket{\bar{T}_0} &=&
\bra{S_g} \left[\ket{T_0} + 
\sum_{i =1}^{4} \frac{\ket{i} \bra{i} H_1 \ket{T_0}}{E(T_0) - E(i)} \right]
\nonumber
\\
&=& 
- \BB_{a\parallel }\bra{S_g} (H'_0)^{-1} \ket{S}.
\eean
Here, $\ket{i}$ and $E(i)$ $(i \in \{1,2,3,4,T_0\})$ are the 
eigenstates and eigenvalues of $H_0$, respectively. 
In the second step, we used the form of $H_1$ given above,
the value $E(T_0) = 0$, 
and the fact that 
$ \sum_{i=1}^{4}  \frac{\ket{i} \bra{i} }{E(i)} 
= (H'_0)^{-1} $.
The $4\times 4$ matrix $H'_0$ can be inverted analytically, yielding
$\langle S_g \ket{\bar{T}_0} = - \frac{\BB_{a\parallel}}{\sqrt 2 t}$.
Importantly, this result is independent of the (1,1)-(0,2)
energy detuning $\Delta$, which is apparent from the above 
definition of $\BB_{a\parallel}$; this $\Delta$-independence 
provides the  feature of charge-noise-resilience 
of the proposed magnetometer,
discussed in Sec.~\ref{sec:noise}.
Using the conditions $B_z \ll B_x$, $\alpha \ll 1$,
and $g_\perp \ll g_\parallel$, fulfilled by the 
realistic parameter
set used in the main text, 
we find
$
\BB_{a\parallel} \approx \mu_B B_z \alpha \frac{g_\parallel^2}{g_\perp}
$,
leading to the result \eqref{eq:overlap}.

\bibliography{sensitivityNotes}

\end{document}